\documentclass[11pt,a4paper]{article}
\usepackage[utf8]{inputenc}
\usepackage[T1]{fontenc}
\usepackage{amsmath,amssymb}
\usepackage{graphicx}
\usepackage[margin=2.5cm]{geometry}
\usepackage[hidelinks]{hyperref}
\usepackage{booktabs}
\usepackage{authblk}
\usepackage{caption}
\usepackage{times}
\usepackage{url}
\usepackage{microtype}

\newcommand{\ket}[1]{\left|#1\right\rangle}
\newcommand{\bra}[1]{\left\langle#1\right|}
\newcommand{\A}{\mathcal{A}}
\newcommand{\good}{\mathrm{good}}
\newcommand{\bad}{\mathrm{bad}}

\title{\bfseries Quantum-walk transport oracles and amplitude estimation\\
of a scalar range statistic in proton therapy}

\author{Kwangzoo Chung}
\affil{Department of Radiation Oncology, Samsung Medical Center, Seoul, Republic of Korea}
\date{}

\begin{document}
\maketitle

\begin{abstract}
\noindent\textbf{Objective.} Quantum amplitude estimation (QAE) promises a quadratic reduction in the sampling cost of Monte Carlo (MC) particle transport, but two obstacles separate it from radiotherapy: the lack of an efficient unitary encoding of charged-particle transport, and the mismatch between QAE's single-scalar output and the per-voxel structure of dose calculation. We address both.

\smallskip\noindent\textbf{Approach.} We construct a discrete-time quantum-walk transport oracle for protons: an energy-dependent coin, a surrogate of the continuous-slowing-down stopping power, drives an energy register, so that energy loss and range straggling follow from the circuit. The oracle is unitary, with a gate count linear in the number of energy levels. The probability that a proton penetrates beyond a distal depth threshold is a directly marked amplitude of its final state, which we estimate by maximum-likelihood amplitude estimation (MLQAE) with Grover iterations and finite shots. A three-gate resource analysis (output structure, amortization, error-corrected wall-clock) delimits where such estimates could outperform classical MC.

\smallskip\noindent\textbf{Main results.} The oracle reproduces a Bragg-shaped depth-deposit profile agreeing with a classical MC of identical rules to within $0.27\%$. Grover amplification matches $\sin^{2}((2k+1)\theta)$ to machine precision up to $k=32$. At equal oracle budget the MLQAE error falls as $N^{-0.93}$, against $N^{-0.49}$ for sampling, and is 4.8 times smaller at the largest budget. Full dose maps are structurally unfavourable irrespective of hardware. The scalar statistic falls short of a wall-clock crossover at clinical scale: the oracle's derived cost, dominated by the multiplexed coin, is $1.0\times10^{6}$ logical gates per call ($3.5\times10^{7}$ with rotation synthesis), 160 to 5400 times the permitted budget.

\smallskip\noindent\textbf{Significance.} Within medical physics, this is to our knowledge the first charged-particle transport oracle with explicit energy dynamics and the first end-to-end amplitude estimation of a transport observable in particle therapy. It derives from the circuit, instead of assuming, a quantitative boundary on where quantum advantage in dose calculation can arise.
\end{abstract}


\section{Introduction}

Monte Carlo (MC) simulation is the reference method for dose calculation in proton and ion-beam therapy, underpinning treatment-planning verification, beam-model commissioning, and research applications from linear-energy-transfer (LET) painting to variable-RBE evaluation. Its accuracy comes at a statistical price: the uncertainty of an MC estimate decreases only as $N^{-1/2}$ in the number of particle histories, so each additional decimal digit of precision costs a hundredfold increase in computation. Decades of work on variance reduction, GPU acceleration, and deep-learning surrogates have attacked the constant in front of this scaling. The exponent itself has not moved.

Quantum amplitude estimation (QAE)~\cite{brassard2002,montanaro2015} changes the exponent. By encoding a target probability as the amplitude of a marked quantum state and amplifying it with Grover iterations, QAE estimates that probability with error scaling as $N^{-1}$ in the number of oracle calls, a quadratic speedup over classical sampling. The scheme has been demonstrated for financial risk integrals~\cite{woerner2019,stamatopoulos2020}, and recently for radiation-adjacent problems: a quantum-walk encoding of photon attenuation in water combined with iterative QAE~\cite{lee2025}, and reactor-physics expectation values estimated with phase-estimation-based QAE~\cite{miao2026}. Speedups for radiotherapy dose estimation have also been reported from linear-solver and variational approaches, with factors of 8--15 attributed to HHL- and VQE-based components~\cite{screp2025}. Without an accompanying resource statement such figures are hard to evaluate. HHL returns a quantum state encoding the solution vector, and its logarithmic scaling is contingent on efficient state preparation and on not requiring classical readout of the full solution~\cite{aaronson2015}. A per-voxel dose map is exactly such a readout. Deterministic quantum solvers for transport-type equations meet the same condition: a quantum lattice-Boltzmann algorithm for radiative transfer claims exponential acceleration, and names the measurements needed to read out the solution and the encoding of the initial state as its open problems~\cite{igarashi2024}. We do not set out to adjudicate any particular claim. Section~\ref{sec:resource} supplies the terms in which claims of this kind can be assessed.

Two obstacles separate these demonstrations from particle-therapy dose calculation. The first is encoding. QAE presupposes a unitary oracle $\A$ that prepares the transport probability distribution. If that distribution has to be loaded generically, state preparation costs $O(2^{n})$ gates for an $n$-qubit register~\cite{shende2006}, a cost that Miao \textit{et al} themselves report as prohibitive beyond nine qubits. There is a more basic difficulty. Generic loading assumes the distribution is already known, and that is what the simulation is supposed to compute. An efficient oracle must therefore generate the distribution from the physics. Discrete-time quantum walks have been proposed as oracles of this kind for radiation transport. Olivier and Nowak encode two-dimensional neutral-particle transport for shielding problems with a position-dependent coin and propose amplifying the walk's amplitudes~\cite{olivier2024}, and the construction of Lee \textit{et al} applies the idea to photon attenuation. Neither carries an energy variable. The first homogenizes the medium into fixed macroscopic cross sections, and the second treats photons in a binary survive-or-absorb model, so neither can represent the defining feature of charged-particle transport: continuous energy loss, range straggling, and the Bragg peak. Energy-resolved quantum encodings of particle evolution have been developed in high-energy physics, notably for parton-shower and detector simulation~\cite{bepari2022,bauer2024}, where the observables of interest are collision-scale and not dosimetric. Adapting that idea to therapeutic proton transport, and to the depth-resolved quantities that dose calculation requires, is the encoding problem we take up here.

The second obstacle is the shape of the clinical question. A treatment-planning dose calculation outputs the dose in each of roughly a million voxels, all filled simultaneously by a single batch of classical histories. QAE, by contrast, estimates one scalar per run: applied voxel-by-voxel it must repeat the entire procedure a million times, and the quadratic gain is consumed by the multiplicity before hardware costs are even considered. Any credible clinical application of QAE must therefore identify estimands that are scalar by nature. Particle therapy offers several. Range and distal-edge quantities, mean doses to small structures, and dose--volume or NTCP summary inputs are all single numbers of direct clinical significance, and range verification in particular is a standing quality-assurance concern specific to proton therapy~\cite{paganetti2012}.

In this work we address both obstacles and quantify what remains. We extend the quantum-walk transport encoding to charged particles by adding an energy register whose stochastic depletion is driven by an energy-dependent coin, a discrete surrogate of the CSDA stopping power. A Bragg-shaped depth-deposit profile, range straggling included, then emerges from the circuit dynamics of a fully unitary oracle whose gate count is linear in the number of energy levels. On this oracle we formulate the probability of penetrating beyond a distal depth threshold as a directly marked amplitude, and we carry out genuine end-to-end amplitude estimation. The Grover operator is applied as an explicit operator sequence, cross-validated against a literal circuit implementation at machine precision, and the amplitude is recovered by maximum-likelihood QAE from finite-shot measurements. It outperforms a matched-budget classical baseline by a margin that grows with the budget.

Finally, we embed these results in a three-gate resource analysis: per-voxel output structure, classical amortization across simultaneous estimands, and wall-clock overhead under quantum error correction. The analysis uses explicit constants and not asymptotics, and it delimits where quantum estimation of transport quantities could and could not overtake classical MC. The gates run from structural to technological, so that a failure at the first cannot be repaired by hardware progress while a failure at the last might be. On this accounting, full dose maps fail structurally. The scalar statistic passes the structural gates but does not reach a wall-clock crossover at clinical scale once the gate count of the oracle is derived instead of assumed. The multiplexed coin that gives the oracle its physics also gives it a cost of order $10^{6}$ logical gates and $4\times10^{5}$ arbitrary-angle rotations per call, two to more than three orders of magnitude above what a crossover at attainable precision would permit. We report the reduction that would be required and identify the term responsible for it.

\section{Background}
\label{sec:background}

\subsection{Amplitude estimation}

Amplitude estimation operates on a unitary $\A$ acting on a register of $n$ qubits that prepares
\begin{equation}
\A\ket{0}^{\otimes n} \;=\; \sqrt{1-a}\,\ket{\psi_{\bad}} \;+\; \sqrt{a}\,\ket{\psi_{\good}},
\label{eq:prep}
\end{equation}
where $\ket{\psi_{\good}}$ is the projection onto a marked (``good'') subspace and $a=\sin^{2}\theta$ is the quantity to be estimated. The Grover iterate
\begin{equation}
Q \;=\; -\,\A\,S_{0}\,\A^{\dagger}\,S_{\chi},
\qquad S_{0}=I-2\ket{0}\!\bra{0},
\qquad S_{\chi}=I-2P_{\good},
\label{eq:grover}
\end{equation}
acts as a rotation by $2\theta$ within the two-dimensional subspace spanned by the good and bad components, so that
\begin{equation}
P_{\good}(k) \;=\; \bigl|\bra{\psi_{\good}}Q^{k}\A\ket{0}\bigr|^{2} \;=\; \sin^{2}\bigl((2k+1)\theta\bigr).
\label{eq:amplification}
\end{equation}
The overall minus sign in~\eqref{eq:grover} places the eigenphases of $Q$ at $\pm 2\theta$; omitting it leaves $P_{\good}(k)$, and with it the MLQAE estimator used below, unchanged. Canonical QAE reads $\theta$ from the eigenphase of $Q$ by quantum phase estimation (QPE), at the cost of an ancillary counting register and a circuit depth of $2^{m}-1$ applications of $Q$. QPE-free variants trade some of this depth for classical post-processing. We use maximum-likelihood amplitude estimation (MLQAE)~\cite{suzuki2020}: $P_{\good}(k)$ is measured with a finite number of shots for a schedule of Grover powers $k$, and $\theta$ is obtained by maximizing the joint likelihood of the observed counts under~\eqref{eq:amplification}. MLQAE requires no counting register and no QPE, and retains the $O(1/N_{\mathrm{oracle}})$ error scaling that constitutes the quadratic advantage over the $O(1/\sqrt{N})$ scaling of classical sampling.

The premise of~\eqref{eq:prep} is an oracle $\A$ that \emph{prepares} the distribution of interest. Loading an arbitrary $n$-qubit amplitude vector requires $O(2^{n})$ entangling gates. It also presupposes knowledge of the distribution that the simulation is meant to produce, and for simulation that is the decisive objection. An oracle that generates the distribution from the transport physics is therefore not an optimization but a precondition.

\subsection{Quantum walks as transport oracles}

A discrete-time quantum walk evolves a walker on a position register under repeated application of a coin operation, which places the coin register in superposition, and a shift operation, which updates the position conditioned on the coin. Markov-chain transport, in which a particle branches into interaction channels at each step with step-dependent probabilities, maps naturally onto this structure: the coin carries the branching probabilities and the shift carries the state update. Lee \textit{et al}~\cite{lee2025} used this correspondence to encode photon attenuation in water, with a two-outcome coin (interact / do not interact) and complete absorption on interaction. Charged-particle transport is different. The particle does not terminate at a single event; it loses energy progressively, so the interaction probability itself evolves with the particle state. Representing that requires an energy register inside the walk, and this is the extension developed below. The step-wise stochastic view of proton transport that the encoding assumes is not peculiar to the quantum setting. Classical jump stochastic differential equation models of the Bragg peak~\cite{crossley2025} have been benchmarked against Geant4 with proton range agreement within 0.2--0.6\,mm~\cite{dean2026}, which supports the formulation and also supplies a classical point of comparison for it.

\subsection{What the walk does and does not use}
\label{sec:nointerference}

A walk of this kind is a coherent encoding of a classical Markov chain, not a source of interference. We state the reason explicitly here, because the distinction determines where the quantum advantage in this paper comes from.

Each transport step consumes a fresh coin qubit that is never reset and never uncomputed within $\A$. Distinct transport histories therefore end in mutually orthogonal coin configurations: the coin register is a permanent record of which branch was taken at every step. Because the histories are perfectly distinguishable, their amplitudes add incoherently, and the probability of finding the walker at a given depth is exactly the probability the corresponding classical Markov chain assigns to it. We verify this numerically: the position distribution obtained from the statevector agrees with the classical chain to $1.1\times10^{-16}$, at the level of floating-point representation. The agreement with a sampled classical Monte Carlo reported in section~\ref{sec:results} is therefore a structural identity and not a coincidence, tested there only to the precision that sampling allows.

This is a design requirement, not a limitation. Amplitude estimation is useful here because $a$ in~\eqref{eq:prep} is the classical probability of the event of interest, and that identification holds only while the histories remain distinguishable. The coin record cannot in any case be discarded cheaply: distinct coin sequences reach the same $(x,E)$, so the coin is not a function of the other registers and no uncomputation recovers $\ket{0}$ without measurement. The price of keeping them is $n_{\mathrm{step}}$ coin qubits, which section~\ref{sec:G} accounts for.

It follows that the oracle on its own confers no advantage. Measuring the position register of $\A\ket{0}$ repeatedly samples the depth distribution and reproduces classical Monte Carlo, error and all, at $O(1/\sqrt{N})$, and with a far larger constant, since each sample costs a full circuit. Everything gained in this paper is gained by amplitude estimation acting on that state, which requires $\A^{\dagger}$ to be available exactly. That requirement, not the walk, is what forces the fully unitary reset-free construction. Seen this way, the construction is a physics-generated instance of the generic scheme of Montanaro~\cite{montanaro2015}, in which amplitude estimation is applied to a reversible implementation of a classical sampler. What is specific here is that the sampler is generated by the transport rules instead of loaded, and what this paper measures is the cost of that instantiation.

The same identity has a consequence for the classical comparator. Because the walk is a classical Markov chain, its depth distribution can also be computed classically and exactly, by propagating the probability mass over $(x,E)$ step by step. In the one-dimensional model of section~\ref{sec:oracle} a history that is still alive has its depth fixed by the step index, so the propagation runs over the $L$ alive energy values alone and costs fewer than $n_{\mathrm{step}}L$ two-branch updates: $1.3\times10^{4}$ for the one-dimensional clinical instance of section~\ref{sec:G}, with no statistical error at all. For the model as built, then, neither sampling nor amplitude estimation is the best classical route to $a$. The comparisons of section~\ref{sec:resource} are made against MC because MC is the method used in practice for the three-dimensional problem that the clinical register sizes stand for. There the phase space of position, direction and energy is large enough that exact propagation becomes a deterministic transport solve, and sampling is the usual choice. Section~\ref{sec:discussion} adds this as a precondition, G0, to the checklist that the three gates define.

\section{A quantum-walk transport oracle for protons}
\label{sec:oracle}

\subsection{Registers and encoding}

\begin{figure}[htbp]
\centering
\includegraphics[width=\textwidth]{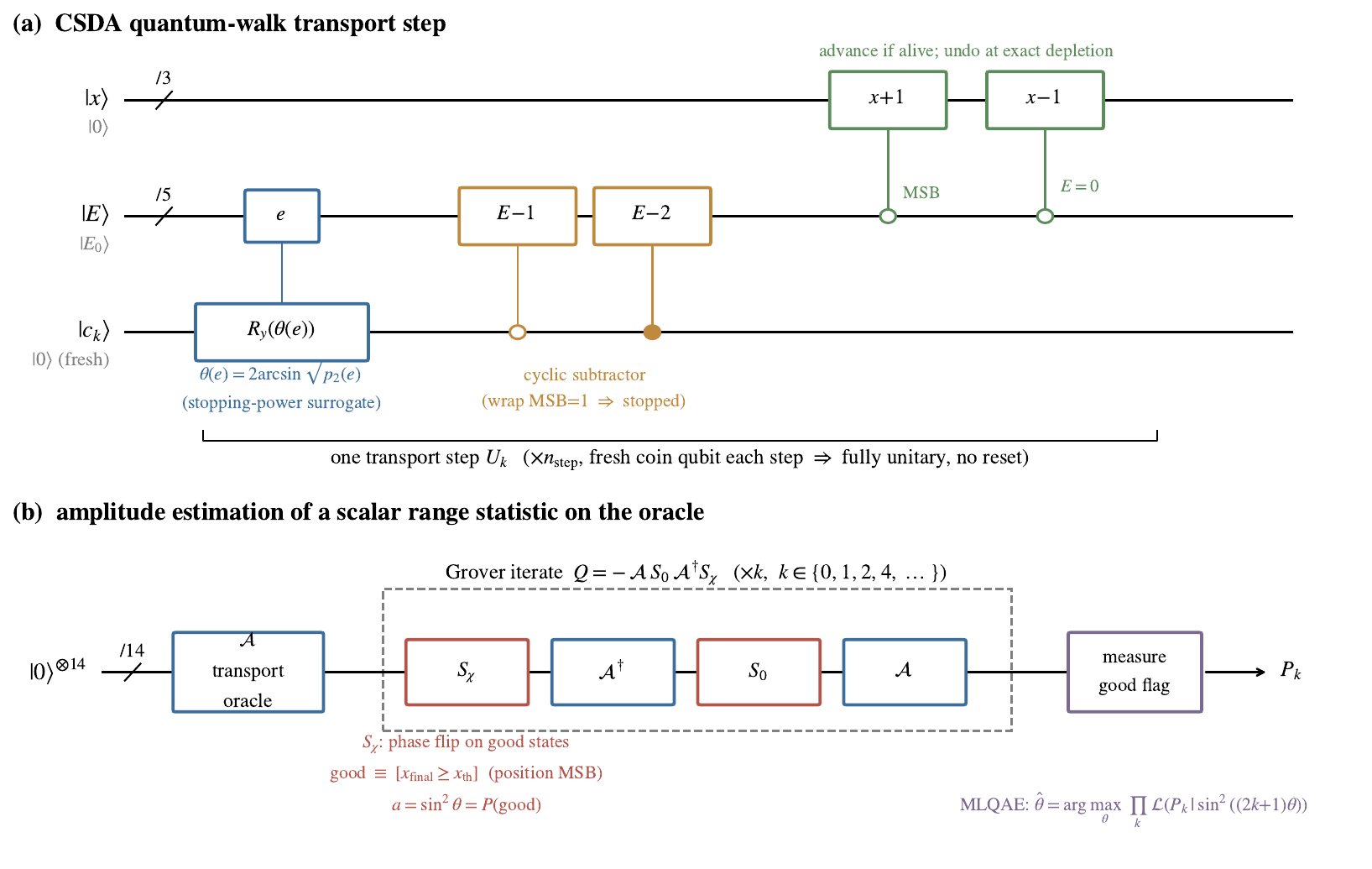}
\caption{Quantum-walk proton transport oracle and amplitude-estimation pipeline. (a) One transport step: an energy-multiplexed coin rotation $R_{y}(\theta(e))$ encoding the stopping surrogate, a coin-controlled cyclic subtractor on the energy register whose wrap flags a stopped particle, and a conditional position advance with correction at exact depletion. A fresh coin qubit is used at each step, so the construction is unitary and requires no mid-circuit reset. (b) Amplitude estimation on the oracle: $\A$ prepares the superposition over histories, the Grover iterate $Q=-\A S_{0}\A^{\dagger}S_{\chi}$ is applied $k$ times, and the good-state flag is measured; $\theta$ is recovered by maximum-likelihood estimation over the schedule of $k$.}
\label{fig:circuit}
\end{figure}

The oracle acts on three registers (figure~\ref{fig:circuit}(a)):
\begin{itemize}
\item a \textbf{position register} of $n_{x}$ qubits holding the depth index $x\in\{0,\dots,2^{n_{x}}-1\}$, initialized to $\ket{0}$;
\item an \textbf{energy register} of $n_{E}$ qubits holding the residual energy $E$ in discrete units, initialized to $\ket{E_{0}}$;
\item a \textbf{coin register} of one fresh qubit $c_{k}$ per transport step, all initialized to $\ket{0}$.
\end{itemize}

A separate coin qubit at each step, instead of a single reused qubit, is what keeps the construction unitary: resetting or measuring an intermediate coin would break the invertibility that $Q=-\A S_{0}\A^{\dagger}S_{\chi}$ requires. The cost is $n_{\mathrm{step}}$ coin qubits; the benefit is that $\A^{\dagger}$ is available exactly (section~\ref{sec:nointerference}).

Depth is discretized in uniform steps and energy in uniform units, with a refinement factor $r$ relating the two: at refinement $r$ the energy unit is $1/r$ of the base unit and the position advances once every $r$ sub-steps, so that in the validation and refinement study an initial energy of 12 base units is $E_{0}=12r$. Refinement converges the energy discretization, but it is not a pure discretization parameter. The loss values of the coin stay at one and two units, so the variance of the energy loss per unit depth falls as $1/r$, and the range straggling falls with it; section~\ref{sec:convergence} quantifies this.

\subsection{Transport step}

One transport step $U_{k}$ consists of three stages.

\paragraph{Stage 1 --- energy-dependent coin.} A multiplexed rotation places the coin in a superposition whose weights encode the local stopping behaviour:
\begin{equation}
U_{\mathrm{coin}}\ket{E=e}\ket{0}_{c} \;=\; \ket{e}\Bigl(\sqrt{1-p_{2}(e)}\,\ket{0}_{c} + \sqrt{p_{2}(e)}\,\ket{1}_{c}\Bigr),
\label{eq:coin}
\end{equation}
implemented as one $n_{E}$-controlled $R_{y}(\theta(e))$ per alive energy value, with
\begin{equation}
\theta(e) = 2\arcsin\sqrt{p_{2}(e)},
\qquad
p_{2}(e) = \min\!\left(0.85,\; \frac{\kappa r}{e}\right).
\label{eq:stopping}
\end{equation}
Here $p_{2}(e)$ is the probability that the step deposits two energy units instead of one, so the mean energy loss per step is $1+p_{2}(e)$ and rises as the residual energy falls. We have kept this surrogate of the continuous-slowing-down (CSDA) stopping power as simple as possible. It reproduces the qualitative $1/E$ rise of the Bethe stopping power in the therapeutic range while keeping the coin a single qubit. In the units used here $\kappa=2.2$, and the cap at $0.85$ prevents the rotation from saturating at very low residual energy; both values were fixed \textit{a priori} and not tuned to any dosimetric target. The two-outcome coin also supplies a discrete form of energy-loss straggling, since histories that draw different coin outcomes accumulate different total losses and stop at different depths. Section~\ref{sec:discussion} discusses replacing it with a tabulated $p(e)$ at no change in circuit structure, and extending the coin to $\log_{2}K$ qubits for $K$ loss channels.

\paragraph{Stage 2 --- energy update.} The coin controls a cyclic subtractor on the energy register:
\begin{equation}
\ket{e}\ket{0}_{c}\;\mapsto\;\ket{e-1}\ket{0}_{c},
\qquad
\ket{e}\ket{1}_{c}\;\mapsto\;\ket{e-2}\ket{1}_{c}.
\label{eq:esub}
\end{equation}
The subtractor is cyclic, so depletion below zero wraps into the upper half of the register. We exploit this: a set most-significant bit of the energy register flags a stopped particle, and the flag is a computational-basis property of the state that the position update reads directly, without any additional arithmetic. Together with the exactly-depleted state $E=0$, the stopped condition is
\begin{equation}
\mathrm{stopped} \iff (E_{\mathrm{MSB}}=1)\ \ \text{or}\ \ (E=0).
\label{eq:stopped}
\end{equation}
The subtractor keeps acting after a particle has stopped. No coin rotation is applied at a flagged or zero energy, so a stopped register is decremented by one unit per step, and its most-significant bit stays set for at least $2^{n_{E}-1}-1$ further steps. The condition $n_{\mathrm{step}}<2^{n_{E}-1}$ therefore guarantees that no stopped history is revived before the walk ends. Every instance in this paper satisfies it, and section~\ref{sec:G} uses it to size the clinical energy register.

\paragraph{Stage 3 --- position update.} The position advances by one bin if the particle is still alive, implemented as an $E_{\mathrm{MSB}}$-controlled cyclic increment on the position register, followed by an $(E=0)$-controlled decrement that undoes the advance in the single case where the particle is depleted exactly at this step. Both are cyclic permutations of the position register.

The full oracle is $\A = U_{n_{\mathrm{step}}-1}\cdots U_{1}U_{0}$ applied to the initialized registers. Per step the implementation uses one $n_{E}$-controlled $R_{y}$ per alive energy value, two coin-controlled modular subtractors on $n_{E}$ qubits, and two controlled modular adders on $n_{x}$ qubits. The gate count is therefore linear in the number of energy levels and in $n_{\mathrm{step}}$, and no step requires an operation whose cost grows with $2^{n_{x}}$. This is the sense in which the oracle avoids the exponential state-preparation cost discussed above. In the reported simulations the modular adders are supplied as permutation unitaries. A hardware-ready decomposition would use standard ripple-carry or QFT-based modular arithmetic~\cite{cuccaro2004,draper2000} with $O(n)$ two-qubit gates per adder, and we quote resource estimates on that basis.

\subsection{Validation against classical Monte Carlo}

The oracle is validated by comparing the depth-resolved deposited energy against a classical MC sampling the \emph{identical} stochastic rules~\eqref{eq:coin}--\eqref{eq:stopped}. The quantum profile is obtained without sampling: at each step the exact joint distribution over $(x,E,c_{k})$ is read from the statevector, and the deposit $\min(1+c,E)$ is accumulated with its probability weight. All circuits are executed as exact statevector simulations with PennyLane~\cite{pennylane}. The classical reference draws $4\times10^{5}$ to $2\times10^{6}$ independent histories with the same $p_{2}(e)$ and the same stopping and advance logic. We assess agreement bin by bin over the depth range carrying non-negligible deposit, against the exact standard error of the per-history deposit in each bin, and also by comparing the deposit summed over the scored depth window. Energy conservation is checked separately and exactly, by verifying that the deposit of a run carried to exhaustion sums to $E_{0}$; that check requires no classical reference.

Two Bragg-peak sharpness metrics are used to follow the profile under refinement of the energy discretization: the ratio of peak to entrance deposit, and the distal 80\%--20\% falloff width obtained by linear interpolation between adjacent bins on the distal side of the peak. Both are evaluated at refinement $r=1,2,4,8,16$, corresponding to 5 to 9 energy qubits, and the resulting peak-to-entrance sequence is compared against the closed-form limit that the transport rules impose on that ratio, derived in section~\ref{sec:results}. The range straggling $\sigma_{R}/R$ is computed alongside them, from the exact stopping-depth distribution of the same rules. Because the number of coin qubits grows with $n_{\mathrm{step}}$, direct statevector simulation of the full circuit is feasible only at $r=1$. The profiles at $r\ge2$ are therefore computed by exact propagation of the identical rules as a classical Markov chain, which by section~\ref{sec:nointerference} is the oracle's own depth distribution. No sampling enters them.

\section{Amplitude estimation of a scalar range statistic}
\label{sec:estimation}

\subsection{Estimand}

Full dose maps are a poor match for amplitude estimation (section~\ref{sec:resource}), so we target a scalar quantity instead. We take the probability that a proton penetrates to or beyond a distal depth threshold $x_{\mathrm{th}}$,
\begin{equation}
a \;=\; P\bigl(x_{\mathrm{final}} \ge x_{\mathrm{th}}\bigr),
\label{eq:estimand}
\end{equation}
where $x_{\mathrm{final}}$ is the depth index at the end of the walk. A history that has stopped keeps its stopping depth, and one still moving has reached depth $n_{\mathrm{step}}$, so for any walk of at least $x_{\mathrm{th}}$ steps $a=P(R\ge x_{\mathrm{th}})$, the survival function of the stopping-depth distribution at $x_{\mathrm{th}}$. In the demonstration geometry $x_{\mathrm{th}}=4$, the distal half of the depth range, and the predicate is the most-significant position bit alone. The quantity is non-decreasing in the initial energy: a shift of the beam's initial energy displaces the stopping distribution and changes $a$ accordingly (section~\ref{sec:limitations}). It is not one of the range descriptors in clinical use. Those are $R_{90}$, $R_{80}$ and $R_{50}$, the depths at which the depth--dose curve falls to a given fraction of its maximum~\cite{paganetti2012}, whereas~\eqref{eq:estimand} is a probability evaluated at a fixed depth, and we make no claim that any clinical workflow currently calls for it. Our claim is narrower: it belongs to the right \emph{class}. It is scalar and not per-voxel, so gate 1 admits it (section~\ref{sec:resource}). It responds to the physical quantity that range verification is concerned with. And it is a computational-basis predicate on the final state of the oracle, so it marks a subspace directly and needs no auxiliary arithmetic; the amplitude already exists in the prepared state and is not computed classically and re-encoded. Placing $x_{\mathrm{th}}$ near the expected stopping depth makes $a$ most sensitive to a range shift, and a range quantile such as $R_{80}$ would correspond to inverting $a$ over $x_{\mathrm{th}}$; section~\ref{sec:discussion} returns to what estimating a dose-weighted quantity would require.

\subsection{Grover operator and MLQAE}

$S_{\chi}$ applies a phase flip on the marked subspace defined by~\eqref{eq:estimand} and $S_{0}$ on the all-zeros state; with the sign convention of~\eqref{eq:grover}, $Q$ rotates by $2\theta$ with $a=\sin^{2}\theta$. We verify the construction in two stages. First, on a reduced instance ($n_{x}=3$, $n_{E}=4$, $n_{\mathrm{step}}=4$; 11 qubits) the Grover iterate is applied in two ways: as a literal circuit, with the reflections realized as diagonal unitaries and $\A^{\dagger}$ as the adjoint of the transport circuit, and as rank-1 vector arithmetic exploiting $\A S_{0}\A^{\dagger}=I-2\ket{\psi_{0}}\!\bra{\psi_{0}}$. The two are required to agree to machine precision at every $k$. Second, on the main instance ($n_{x}=3$, $n_{E}=5$, $n_{\mathrm{step}}=6$, $E_{0}=7$; 14 qubits) the validated vector form is used, and the amplification law~\eqref{eq:amplification} is checked against state-derived probabilities up to $k=32$. The vector form is used for the estimation runs because materializing the reflection unitaries at 14 qubits is prohibitively expensive in simulator memory, not because the circuit is unavailable; Miao \textit{et al}~\cite{miao2026} report the same reduction for $n\ge9$.

MLQAE uses the Grover schedule $k\in\{0,1,2,4,\dots,k_{\max}\}$ with 100 shots per schedule point. Counts $h_{k}$ are drawn from $\mathrm{Binomial}(100, P_{\good}(k))$ with $P_{\good}(k)$ evaluated from the amplified states, and $\theta$ is estimated as
\begin{equation}
\hat\theta = \arg\max_{\theta}\ \sum_{k}\Bigl[h_{k}\log p_{k}(\theta) + (100-h_{k})\log\bigl(1-p_{k}(\theta)\bigr)\Bigr],
\qquad p_{k}(\theta)=\sin^{2}\bigl((2k+1)\theta\bigr),
\label{eq:mlqae}
\end{equation}
maximized on a uniform grid of $4\times10^{4}$ points in $(0,\pi/2)$. The reported estimate is $\hat a=\sin^{2}\hat\theta$. The total oracle budget of a run is $N_{q}=\sum_{k}100(2k+1)$, counting each application of $\A$ within $Q^{k}\A$, so that $\A$ and its adjoint are charged separately.

The shot model is exact-circuit sampling: no circuit noise, gate infidelity or state-preparation error is included, so the estimation results characterize the estimator and the encoding rather than any hardware (section~\ref{sec:limitations}).

\subsection{Classical baseline}

Because the estimand~\eqref{eq:estimand} is a Bernoulli event at the level of a single history, the classical estimator is unambiguous: draw $N$ histories, count those penetrating to $x_{\mathrm{th}}$, and report the sample proportion, with error $\sqrt{a(1-a)/N}$. Quantum and classical are compared at \emph{equal budget}, $N=N_{q}$, so that one oracle call is charged the same as one classical history. At the level of counts this favours neither side. The additional per-call cost of a fault-tolerant oracle is accounted for separately in section~\ref{sec:resource}. Quantum errors are reported as the median absolute error over 40 independent MLQAE trials, classical errors as the median over $4\times10^{3}$ binomial draws, each with a 95\% bootstrap interval from $2\times10^{3}$ resamples; the same resamples give intervals for the error ratios and the fitted slopes.

\subsection{Robustness checks}

Two checks bound how much of the reported comparison depends on the particular case studied. First, we compare the classical baseline against its analytic form. The median absolute error of a binomial proportion is $0.6745\sqrt{a(1-a)/N}$, and the simulated baseline should reproduce this, apart from the discreteness of the binomial at the smallest budget. That check rules out a comparator weakened by an implementation error. Second, the estimation is repeated at $a=0.05$, $0.10$, $0.25$, $0.50$ and $0.90$ at a fixed budget, using the amplification law~\eqref{eq:amplification} to generate the shot counts, so that only the value of the estimand changes while the schedule, the shot allocation and the likelihood grid are held fixed. Since $\sqrt{a(1-a)}$ is maximized at $a=1/2$, an estimand near that value gives the classical baseline its largest error, so the value arising in the present geometry, $a=0.667$, is not a neutral choice but one that flatters the comparison; the scan is what establishes how much of the reported ratio it carries.

Separately, the response of the estimand to a change in range is measured by varying the initial energy $E_{0}$ by one unit either side of its nominal value, with $n_{\mathrm{step}}$ and $x_{\mathrm{th}}$ held fixed, and recording the resulting change in the penetration probability. This quantifies how strongly the chosen scalar tracks the quantity that range verification is concerned with.

\section{Resource analysis}
\label{sec:resource}

Sections~\ref{sec:oracle} and~\ref{sec:estimation} establish that a scalar range amplitude can be prepared and estimated on a physics-generated oracle with the expected $O(1/N)$ error scaling. Whether that scaling can translate into a clinical benefit is a separate question. It is settled by constants and by the structure of the task, not by asymptotics. We formalize it as three successive gates, each of which a candidate application must pass. Throughout, the failure probability of the estimator is fixed at $\alpha=0.05$ and the query cost of amplitude estimation is taken from the iterative-QAE bound~\cite{grinko2021},
\begin{equation}
N_{\mathrm{QAE}}(\epsilon) \;=\; \frac{6}{\epsilon}\,\ln\!\left(\frac{2}{\alpha}\log_{2}\frac{\pi}{4\epsilon}\right),
\label{eq:iqae}
\end{equation}
which includes the shot overhead that asymptotic statements omit. The constant 6 is the worst case that Grinko \textit{et al} observed empirically for the Chernoff--Hoeffding variant of iterative QAE; the bound they prove for that variant has the constant 50~\cite{grinko2021}. We use~\eqref{eq:iqae} rather than the MLQAE schedule of section~\ref{sec:estimation} because it is published and explicit; the two differ by $O(1)$ factors that do not affect the conclusions. The classical comparator is the sample proportion, requiring $N_{\mathrm{MC}}(\epsilon)=\sigma^{2}/\epsilon^{2}$ histories, with $\sigma^{2}=a(1-a)\le 1/4$ taken at its worst case.

The two sides carry different confidence conventions. Equation~\eqref{eq:iqae} guarantees its accuracy with probability $1-\alpha$, whereas $\sigma^{2}/\epsilon^{2}$ is a $1\sigma$ statement. Matching the classical count to the same $95\%$ interval multiplies it by $z_{0.975}^{2}=3.84$ in favour of the quantum side: the gate-1 ratio falls from 700 to 181, the gate-2 unity crossing moves from $\epsilon=6.9\times10^{-4}$ to $2.8\times10^{-3}$, and the gate-3 crossovers move from $5.9\times10^{-9}$ and $6.0\times10^{-8}$ to $2.3\times10^{-8}$ and $2.3\times10^{-7}$ at $t_{\mathrm{logical}}=1\,\mu$s and $0.1\,\mu$s. Replacing the empirical constant 6 in~\eqref{eq:iqae} by the proven 50 multiplies every quantum count by 8.3 in the other direction: the gate-1 ratio rises to $5.8\times10^{3}$, the gate-2 crossing moves to $\epsilon=8.0\times10^{-5}$, and the gate-3 crossover at $0.1\,\mu$s to $7.0\times10^{-9}$. No verdict changes under either substitution. We quote the $1\sigma$ classical count, the convention in which MC dose-calculation uncertainties are reported, and the empirical constant, which reflects the query counts that iterative QAE needs in practice.

All three gates take sampling as the classical method. They are therefore a test for the three-dimensional problem that the clinical register sizes represent, and they presuppose the check listed as G0 in section~\ref{sec:discussion}: that the estimand cannot be obtained without sampling by the best available classical method, as it can for the one-dimensional model as built (section~\ref{sec:nointerference}).

\subsection{Gate 1 --- output structure}

Classical MC fills every voxel of the dose grid from a single batch of histories: each history deposits along its whole track, so the cost of the calculation is independent of how many voxels are scored. Amplitude estimation has no such property. Each voxel dose is a distinct amplitude requiring its own oracle, its own Grover schedule, and its own estimation run, so the quantum cost of a dose map is multiplied by the number of scored voxels $V$:
\begin{equation}
N_{\mathrm{quantum}}^{\mathrm{map}} = V\,N_{\mathrm{QAE}}(\epsilon),
\qquad
N_{\mathrm{classical}}^{\mathrm{map}} = N_{\mathrm{MC}}(\epsilon).
\label{eq:gate1}
\end{equation}
We evaluate~\eqref{eq:gate1} for a clinical grid of $128^{3}=2^{21}$ voxels and, more favourably, for the 2\% of voxels carrying high dose, at a target of $\epsilon=10^{-2}$. The classical anchor is taken from a measured GPU proton dose engine: gPMC reaches approximately 1\% relative statistical uncertainty with $10^{7}$ source protons in 6--22\,s~\cite{jia2012}. We report gate 1 against this measured value and, as an alternative, against a ten-fold more demanding $10^{8}$-history anchor, so that the conclusion can be read either way. Equation~\eqref{eq:gate1} charges each voxel at an absolute accuracy $\epsilon$, whereas the classical anchor is a relative uncertainty. For a voxel receiving a fraction $f$ of the deposit, a relative accuracy of 1\% needs $\epsilon=0.01f$, so the comparison is lenient to the quantum side by about $1/f$. No oracle cost or clock model enters gate 1. It compares call counts and nothing else, which makes it the least model-dependent of the three gates and the one a candidate application has to pass first.

\subsection{Gate 2 --- amortization across estimands}

A treatment plan is characterized not by one scalar but by a set of them: several DVH points, mean doses to a handful of organs at risk, distal-range quantities, NTCP inputs. Classical MC again amortizes, since the same batch of histories scores all $K$ quantities, whereas amplitude estimation repeats. Comparing at a target $\epsilon$,
\begin{equation}
N_{\mathrm{quantum}}(K) = K\,N_{\mathrm{QAE}}(\epsilon),
\qquad
N_{\mathrm{classical}}(K) = N_{\mathrm{MC}}(\epsilon),
\label{eq:gate2}
\end{equation}
so the naive quadratic advantage is reduced by a factor $K$. We evaluate~\eqref{eq:gate2} for $K=10$ at $\epsilon=10^{-2}$, $10^{-3}$ and $10^{-4}$. Gate 2 determines whether the scalar-estimand strategy that gate 1 motivates is itself viable.

\subsection{Gate 3 --- wall-clock under error correction}

Gates 1 and 2 count oracle calls, charging one call the same as one classical history. That equivalence is generous. A fault-tolerant oracle call is a circuit of many logical gates, and each of those is itself an error-corrected operation with substantial overhead~\cite{fowler2012}. Gate 3 converts both sides to wall-clock time,
\begin{equation}
t_{\mathrm{quantum}} = N_{\mathrm{QAE}}(\epsilon)\,G\,t_{\mathrm{logical}},
\qquad
t_{\mathrm{classical}} = N_{\mathrm{MC}}(\epsilon)/R,
\label{eq:gate3}
\end{equation}
where $G$ is the number of logical gates per oracle call, $t_{\mathrm{logical}}$ the duration of a logical gate, and $R$ the classical throughput. We take $R=10^{6}$ histories per second, which sits within the $4.5\times10^{5}$--$1.7\times10^{6}$ range implied by gPMC timings~\cite{jia2012} and is conservative relative to more recent GPU engines, and evaluate three logical-gate times: $t_{\mathrm{logical}}=10\,\mu$s, corresponding to a surface-code cycle of about $1\,\mu$s at a code distance around 20; $1\,\mu$s; and $0.1\,\mu$s, the last being roughly two orders of magnitude beyond current superconducting performance and included as an upper bound on plausible improvement. We do not assume $G$. We derive it from the circuit of section~\ref{sec:oracle} and compare it against generic amplitude loading, for which $G=2^{22}$ is the state-preparation cost of the 21-qubit register that a $128^{3}$ grid requires. The crossover accuracy is the value of $\epsilon$ at which $t_{\mathrm{quantum}}/t_{\mathrm{classical}}$ falls below unity.

\subsection{Derived gate count of the transport oracle}
\label{sec:G}

The oracle of section~\ref{sec:oracle} is efficient in the asymptotic sense: its cost is linear in the number of steps and of energy levels, and contains no term in $2^{n_{x}}$. For~\eqref{eq:gate3}, though, the constants are what matter, so we count them, as two-qubit-equivalent logical gates with each single-qubit rotation counted as one gate. An $n$-controlled rotation implemented with an ancilla-assisted V-chain costs approximately $12(n-1)+4$ gates~\cite{barenco1995}, a ripple-carry modular addition of a constant on $m$ qubits approximately $4m$ gates~\cite{cuccaro2004}, and its controlled form approximately twice that plus the control chain. One transport step then costs
\begin{equation}
g_{\mathrm{step}} \;=\; g_{\mathrm{coin}} \;+\; 2\,c(n_{E}) \;+\; c(n_{x}) \;+\; c_{n_{E}}(n_{x}),
\label{eq:gatecount}
\end{equation}
where $c(\cdot)$ denotes the controlled-adder costs above. The multiplexed coin can be built in two ways. Implemented literally, as one $n_{E}$-controlled rotation for each of the $L$ energy values at which a rotation is applied, it costs $g_{\mathrm{coin}}=L[12(n_{E}-1)+4]$; this is how the demonstration circuit is written. As a uniformly controlled rotation~\cite{mottonen2004} it costs $2^{n_{E}}$ CNOTs and $2^{n_{E}}$ single-qubit rotations, $g_{\mathrm{coin}}=2^{n_{E}+1}$, independent of $L$. We adopt the second for all resource estimates, since it is the standard synthesis and the cheaper one at every size considered here. In both forms the coin carries every arbitrary-angle rotation in the step, $2^{n_{E}}$ of them for the uniformly controlled form, and those rotations, not the CNOTs, are what error correction charges for.

For the 14-qubit demonstration instance ($n_{x}=3$, $n_{E}=5$, $n_{\mathrm{step}}=6$) the circuit as implemented applies the coin rotation at $L=12$ energy values, of which the seven reachable from $E_{0}=7$ are the minimum required. This gives $g_{\mathrm{step}}\approx800$ and $G=g_{\mathrm{step}}n_{\mathrm{step}}\approx4.8\times10^{3}$ for the circuit as written ($3.2\times10^{3}$ at $L=7$, and $1.4\times10^{3}$ with a uniformly controlled coin), with a Grover iterate costing $G_{Q}=2G+2\times(\text{reflection cost})\approx9.9\times10^{3}$.

The clinical register sizes follow from three conditions. The position register indexes a $128^{3}$ grid, $n_{x}=21$, as a three-dimensional extension of the walk would require, and $n_{\mathrm{step}}=384=3\times128$ is the length of a lattice path across it. The one-dimensional model needs neither; both are kept so that the estimate refers to the geometry of a clinical calculation. The energy register must hold an initial energy whose range spans the walk, and $E_{0}\approx400$ places the mean stopping step at 390. Only values below the stop flag are alive, so $E_{0}\le2^{n_{E}-1}-1$, and the flag must survive the rest of the walk, so $n_{\mathrm{step}}<2^{n_{E}-1}$ (section~\ref{sec:oracle}). Both conditions require $n_{E}=10$. With the coin as a uniformly controlled rotation these give
\begin{equation}
g_{\mathrm{step}}\approx2.7\times10^{3},
\qquad
G\approx1.0\times10^{6},
\qquad
G_{Q}\approx2.0\times10^{6},
\label{eq:Gclinical}
\end{equation}
on 415 qubits, of which 384 are coin qubits, with $3.9\times10^{5}$ arbitrary-angle rotations per call. The literal coin, with a rotation at each of the $L=E_{0}=400$ reachable energies, would give $G\approx1.7\times10^{7}$. Generic loading of the 21-qubit position register, at $2^{22}=4.2\times10^{6}$ gates and $2^{21}$ rotations, has a $T$-equivalent count of $2.0\times10^{8}$ on the same accounting. The generated oracle is therefore cheaper by factors of 4 and 6, less than an order of magnitude at this scale, because the exponential cost has moved from $2^{n_{x}}$ to $n_{\mathrm{step}}2^{n_{E}}$ and has not disappeared. The case for generating the distribution rests on the argument of section~\ref{sec:background}, that loading presupposes the answer, and not on the gate count. The exact stopping-depth distribution of this instance has $\sigma_{R}/R=0.7\%$, of the same order as the physical straggling of protons~\cite{bortfeld1997}, so no further energy refinement is charged (section~\ref{sec:convergence}). A one-dimensional depth-only calculation at 2\,mm resolution over a 30\,cm range ($n_{x}=8$, $n_{\mathrm{step}}=150$, $E_{0}\approx160$, $n_{E}=9$) gives $G\approx2.1\times10^{5}$.

These figures are one to two orders of magnitude above the $10^{4}$ that a naive reading of ``efficient oracle'' might suggest, and they change the conclusion of gate 3. With $G\approx1.0\times10^{6}$, no crossover occurs above $\epsilon=10^{-9}$ at $t_{\mathrm{logical}}=10\,\mu$s; at $1\,\mu$s it falls at $\epsilon\approx6\times10^{-9}$, and at the aggressive $0.1\,\mu$s at $\epsilon\approx6\times10^{-8}$, a precision at which the classical calculation itself would take about $2\times10^{4}$ hours. Gate 3 is therefore not passed at clinical scale by the oracle as constructed. On balance the estimate is optimistic. In its favour, the synthesis costs we assume are textbook and not optimized. Against it, \eqref{eq:gatecount} counts logical gates, whereas the binding cost of an error-corrected operation is the $T$ count. Synthesizing each of the $3.9\times10^{5}$ rotations per call to a total error of $10^{-3}$ takes about 86 $T$ gates~\cite{ross2016}, a $T$-equivalent count of $3.5\times10^{7}$ per call that moves the $0.1\,\mu$s crossover to $\epsilon\approx1.7\times10^{-9}$. Against it also, and more importantly, the counts in~\eqref{eq:Gclinical} are those of the \emph{present} physics model: a two-outcome coin in one dimension. A model adequate for dosimetry would need a multi-channel coin and lateral transport, and both increase $g_{\mathrm{step}}$. Making the physics more realistic would therefore reinforce the gate-3 conclusion, not overturn it. Section~\ref{sec:discussion} examines which terms in~\eqref{eq:gatecount} are reducible and what reduction would be required.

\subsection{Scope and known assumptions}

Gate 1 depends on no hardware parameter, gate 2 on none beyond $K$, and only gate 3 introduces quantities whose future values are unknown. Five parameters enter the analysis: the $10^{7}$- and $10^{8}$-history classical anchors, $R=10^{6}$ histories per second, the logical-gate times, the $1\sigma$ convention for the classical count, and the constant in~\eqref{eq:iqae}. We state them here so that a reader can substitute institution-specific measurements.

\section{Results}
\label{sec:results}

\subsection{The oracle reproduces the intended transport model}
\label{sec:validation}

\begin{figure}[htbp]
\centering
\includegraphics[width=0.72\textwidth]{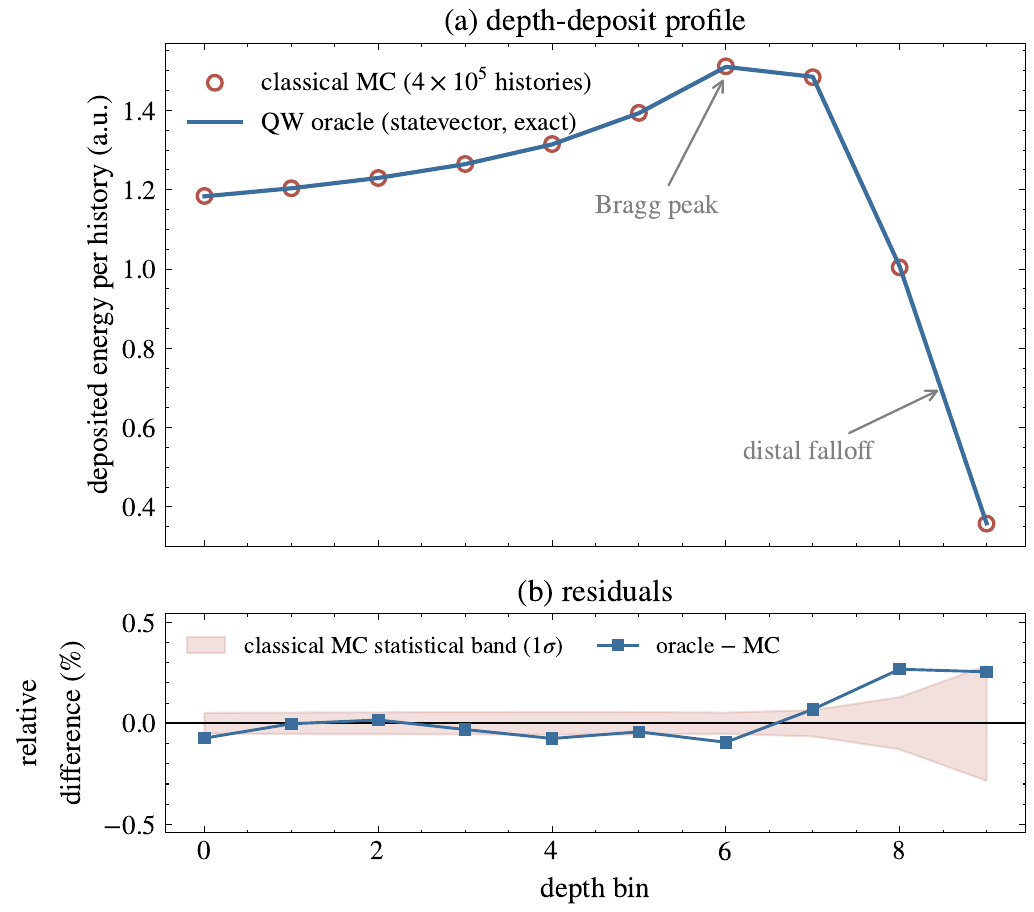}
\caption{Transport oracle validated against a classical Monte Carlo of identical stochastic rules. Validation instance: $n_{x}=4$, $n_{E}=5$, $n_{\mathrm{step}}=10$, $E_{0}=12$ at $r=1$. (a) Depth-resolved deposited energy: the oracle profile is exact, read from the statevector, while the classical reference uses $4\times10^{5}$ histories. (b) Relative residuals, with the classical $1\sigma$ sampling band shaded; the band is the exact standard error of the per-history deposit in each bin, against which $|z|\le2.1$. Agreement is within 0.267\% in every bin carrying appreciable deposit, and the deposit summed over the scored window agrees to $4\times10^{-5}$ relative; energy conservation is verified separately and exactly (section~\ref{sec:validation}).}
\label{fig:validation}
\end{figure}

Figure~\ref{fig:validation} compares the depth-resolved deposited energy produced by the transport oracle against a classical Monte Carlo sampling the identical stochastic rules. The validation instance is $n_{x}=4$, $n_{E}=5$, $n_{\mathrm{step}}=10$ at refinement $r=1$, so that $E_{0}=12$ by the convention of section~\ref{sec:oracle}; it is deeper and higher in initial energy than the estimation instance of section~\ref{sec:estimation}, which uses $n_{x}=3$ and $E_{0}=7$. The oracle profile is exact, read from the statevector and not sampled, while the classical reference uses $4\times10^{5}$ histories. The two agree to within 0.267\% in every depth bin carrying appreciable deposit, with the largest residuals in the two most distal bins where the classical statistical error is itself largest; residuals in the plateau bins are below 0.1\%. Measured against the exact standard error of the per-history deposit in each bin, the residuals have $|z|\le2.1$ and $\chi^{2}/\mathrm{dof}=1.4$ over the ten bins, treating the bins as independent, so they are consistent with the classical sampling error and not systematic. The deposit summed over the scored window agrees to $4\times10^{-5}$ in relative terms (11.9490 versus 11.9485 in energy-register units). Energy conservation is established separately and exactly: carried to exhaustion, the deposit summed over all depth bins equals $E_{0}=12$ to machine precision, which is what confirms that the cyclic arithmetic loses no energy and that the stopping condition~\eqref{eq:stopped} loses no histories. The ten-step window of figure~\ref{fig:validation} carries $99.58\%$ of that total; the remaining $0.42\%$ is deposited by the longest-ranging histories beyond the plotted bins, and the two sums quoted above are therefore window sums, not totals. By section~\ref{sec:nointerference} exact agreement is expected; what figure~\ref{fig:validation} tests is that the implementation realizes the intended process, to the precision the classical sampling allows.

The profile itself has the qualitative structure of a proton depth-dose curve: a slowly rising plateau, a peak at bin 6, and a sharp distal falloff to near zero within two bins. The shape follows from the rules we chose, but it is not fitted and not imposed bin by bin. The whole of the physics input is one scalar function, the energy dependence of the coin angle in equation~\eqref{eq:stopping}. The plateau rise, the peak and the finite width of the falloff all follow from the step structure. The falloff width is the discrete counterpart of range straggling, and it is what distinguishes an energy-resolved walk from the binary survive-or-absorb encoding used for photons.

\subsection{Sharpness and straggling under energy refinement}
\label{sec:convergence}

\begin{figure}[htbp]
\centering
\includegraphics[width=\textwidth]{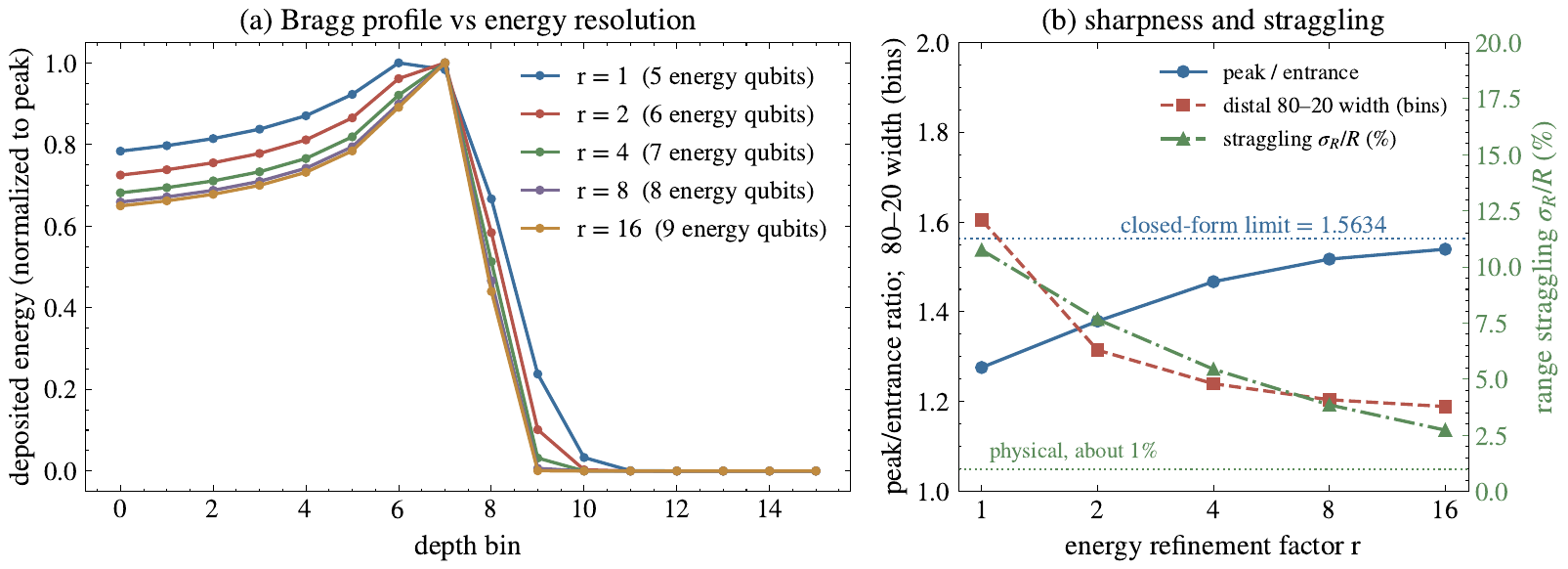}
\caption{Bragg-peak sharpness and range straggling under energy-register refinement. (a) Depth-deposit profiles at refinement $r=1,2,4,8,16$, corresponding to 5 to 9 energy qubits, each normalized to its peak. (b) Peak-to-entrance ratio and distal 80\%--20\% falloff width against $r$ (left axis), and range straggling $\sigma_{R}/R$ from the exact stopping-depth distribution at sub-step resolution (right axis). The peak-to-entrance ratio approaches the closed-form, straggling-free limit $(1+0.85)/(1+\kappa/12)=1.5634$ of equation~\eqref{eq:sharplimit}, well below the value of roughly 3 to 5 seen in measured curves; no coin with losses of one and two units can exceed the bound~\eqref{eq:sharpbound}. The straggling falls as $10.9\%/\sqrt{r}$ and stays above the physical value of about $1\%$ throughout. All profiles are computed by exact propagation of the identical rules, which is the oracle's own depth distribution (section~\ref{sec:nointerference}).}
\label{fig:resolution}
\end{figure}

The demonstration instance uses a coarse energy register, and its peak-to-entrance ratio of 1.28 is well below the value of roughly 3 to 5 seen in measured proton depth-dose curves. Figure~\ref{fig:resolution} shows how much of this is set by the discretization. Refining the energy unit by a factor $r=1,2,4,8,16$, which corresponds to 5 through 9 energy qubits, raises the peak-to-entrance ratio monotonically from 1.276 to 1.539 and narrows the distal 80\%--20\% falloff from 1.60 to 1.19 depth bins. The successive increments in the peak-to-entrance ratio are $+0.103$, $+0.088$, $+0.050$ and $+0.022$, and the sequence converges to a limit that the transport rules fix in closed form, so there is nothing here to extrapolate. The mean deposit per step is $1+p_{2}(e)$, equal to $1+\kappa/12=1.1833$ at entrance, where $E=E_{0}=12r$, and at most $1+0.85$ wherever the cap in~\eqref{eq:stopping} is active, so
\begin{equation}
\lim_{r\to\infty}\frac{D_{\mathrm{peak}}}{D_{\mathrm{entrance}}}\;=\;\frac{1+0.85}{1+\kappa/12}\;=\;1.5634 .
\label{eq:sharplimit}
\end{equation}
The computed sequence approaches this value from below: the ratio 1.539 at $r=16$ is $98.5\%$ of the limit.

The refinement does more than resolve the energy axis, and the limit~\eqref{eq:sharplimit} should be read with that in mind. The coin's loss values stay at one and two units of $1/r$, so the variance of the energy loss per unit depth falls as $1/r$ and the range straggling falls with it. From the exact stopping-depth distribution of the rules, $\sigma_{R}/R$ is $10.8\%$ at $r=1$, $5.4\%$ at $r=4$ and $2.7\%$ at $r=16$, following $10.9\%/\sqrt{r}$ closely (figure~\ref{fig:resolution}(b)). The physical value for protons in water is about $1\%$~\cite{bortfeld1997}, which this sequence would reach only near $r\approx120$. Every profile in figure~\ref{fig:resolution} therefore carries more straggling than a physical beam, and the narrowing of the distal falloff with $r$ reflects the removal of that excess. Equation~\eqref{eq:sharplimit} is the straggling-free limit of the rules. A refinement that converged to a fixed physical model would hold both the mean and the variance of the loss per unit depth constant, which with two outcomes means rescaling the loss values with $r$; we leave that to the multi-channel coin of section~\ref{sec:discussion}. The clinical-scale instance of section~\ref{sec:G} is not affected in the same way: with $E_{0}\approx400$ at $r=1$, its exact $\sigma_{R}/R$ is $0.7\%$, of the same order as the physical value.

The limit~\eqref{eq:sharplimit} is far below the ratio of roughly 3 to 5 seen in measured proton depth-dose curves, so the residual gap is not closed by refinement alone, and we do not claim that it is. Its origin lies in the loss values of the coin, not in the resolution of the energy register, and this too can be stated as a bound. Because a step deposits either one or two energy units, the mean deposit is below 2 everywhere and equal to $1+\kappa/12$ at entrance, so
\begin{equation}
\frac{D_{\mathrm{peak}}}{D_{\mathrm{entrance}}}\;<\;\frac{2}{1+\kappa/12}\;=\;1.690
\label{eq:sharpbound}
\end{equation}
for every choice of the cap at fixed $\kappa$, and below 2 for every choice of $\kappa$ as well. The bound belongs to the loss set $\{1,2\}$ and not to two-outcome coins in general. With losses $\ell_{1}<\ell_{2}$ the corresponding bound is $\ell_{2}/\mathbb{E}[\ell\,|\,\text{entrance}]$, and a larger second loss raises it: a coin with losses $\{1,4\}$ and the same $p_{2}(e)$ reaches a peak-to-entrance ratio of 1.71 at $r=8$, and its straggling-free limit is 2.29. The price is paid in variance. With the loss values fixed across energies, $p_{2}(e)$ sets the mean loss at each energy and the per-step variance $p_{2}(1-p_{2})(\ell_{2}-\ell_{1})^{2}$ follows from it, so a larger $\ell_{2}$ buys peak height with additional straggling, and a two-outcome coin cannot set the two independently. Matching the stopping power and the straggling of a physical beam together needs more loss channels, at $\log_{2}K$ coin qubits per step, along with the lateral transport and nuclear processes that the one-dimensional model omits. Section~\ref{sec:discussion} discusses both.

\subsection{The range amplitude is genuinely encoded and amplified}

The estimand is the probability that a proton penetrates to the distal threshold, equation~\eqref{eq:estimand}, which for the main instance takes the value $a=0.6674$. Two checks establish that this number is an amplitude of the prepared state rather than a classically computed quantity.

First, on the 11-qubit reduced instance we applied the Grover iterate as a literal circuit, with the reflections realized as diagonal unitaries and $\A^{\dagger}$ as the adjoint of the transport circuit. Over four successive iterations it agrees with the rank-1 vector form to a maximum deviation of $5.5\times10^{-16}$, at the level of numerical precision. The vector form used for the main runs is therefore the circuit, not a substitute for it.

Second, on the 14-qubit main instance the probability of the marked subspace, evaluated from the amplified states themselves, follows $\sin^{2}((2k+1)\theta)$ to within $6\times10^{-15}$ for every $k$ from 0 to 32, reproducing the full oscillation of the amplification law: $P_{\good}$ takes the values 0.6674, 0.0728 and 0.9954 at $k=0$, 1 and 2, and keeps oscillating as the rotation carries the state repeatedly through the marked subspace.

\subsection{End-to-end estimation outperforms the classical baseline}
\label{sec:mlqae-results}

\begin{figure}[htbp]
\centering
\includegraphics[width=\textwidth]{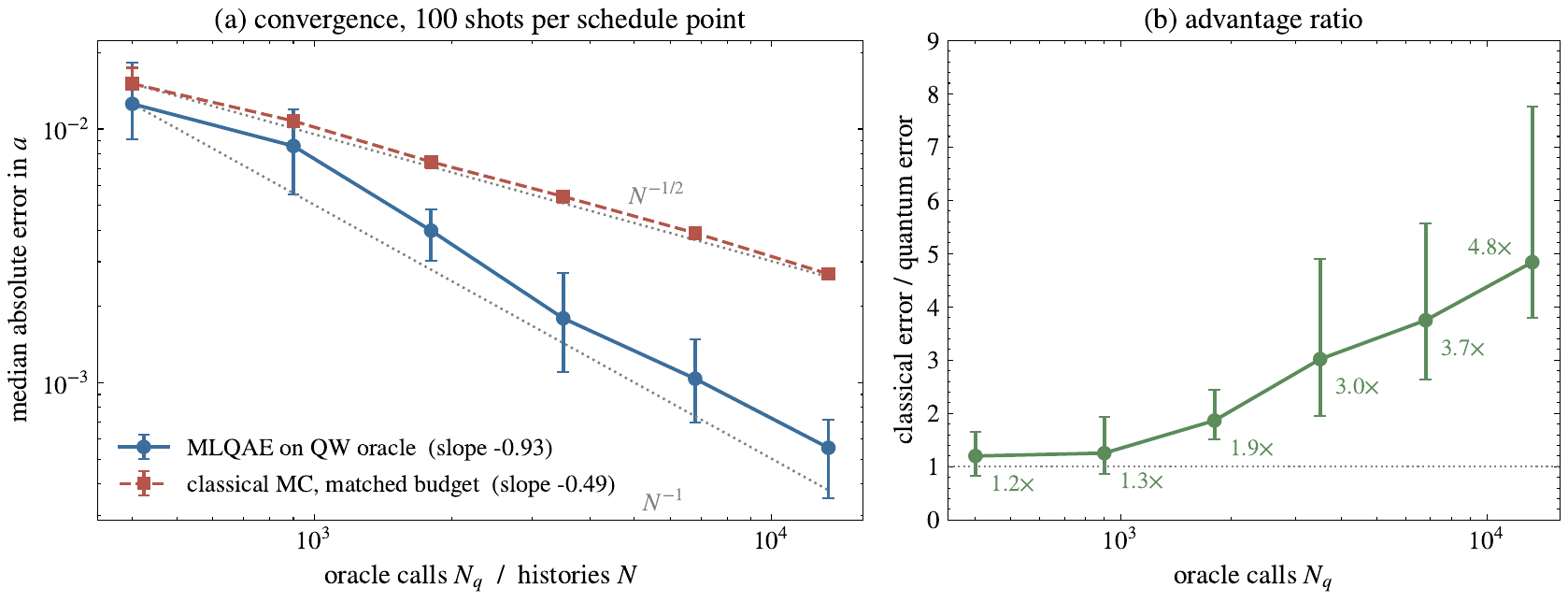}
\caption{End-to-end maximum-likelihood amplitude estimation of the distal penetration probability $a=0.667$ on the transport oracle. (a) Median absolute error over 40 trials against oracle calls, with 95\% bootstrap intervals, with 100 shots per schedule point, compared with a classical Bernoulli estimator at matched budget; dotted lines show $N^{-1}$ and $N^{-1/2}$ references. (b) Ratio of classical to quantum error, with 95\% bootstrap intervals. The comparison is in oracle calls, not in time: the per-call circuit cost is treated in section~\ref{sec:gates-passed}, and these ratios should not be read as speedups.}
\label{fig:convergence}
\end{figure}

Figure~\ref{fig:convergence} reports maximum-likelihood amplitude estimation of $a$ from finite-shot measurements, 100 shots at each point of the Grover schedule, with the median absolute error taken over 40 independent trials. Quantum and classical are compared at equal budget, one oracle call charged as one classical history. The comparison is therefore in oracle calls and not in time: the per-call cost of the circuit is the subject of section~\ref{sec:gates-passed}, and the ratios in figure~\ref{fig:convergence}(b) should not be read as speedups.

The error decreases with a fitted log--log slope of $-0.93$ (95\% interval $-1.07$ to $-0.82$) against $-0.49$ ($-0.52$ to $-0.48$) for the classical baseline, consistent with the $N^{-1}$ and $N^{-1/2}$ scalings expected of amplitude estimation and of sampling respectively. Because the two slopes differ, the advantage is not a constant factor; it grows with budget. The classical-to-quantum error ratio is $1.2\times$ at 400 oracle calls, where its interval (0.8 to 1.7) still includes unity, and rises to $4.8\times$ (3.8 to 7.8) at $1.33\times10^{4}$ calls, where the quantum estimator reaches a median error of $5.5\times10^{-4}$ and the classical estimator is still at $2.7\times10^{-3}$.

Two checks bound how much of this depends on the particular case. The classical baseline reproduces the binomial median error $0.6745\sqrt{a(1-a)/N}$ to within 3\% at every budget from 900 calls, and at 400 calls it is 5\% below it, on the side of a stronger comparator, so it is not a weakened one. Repeating the estimation at $a=0.05$, $0.10$, $0.25$, $0.50$ and $0.90$, and at the value of the present geometry, at a fixed budget of $6.8\times10^{3}$ calls gives ratios between 4.2 and 6.3, a spread of less than a factor of 1.5. The advantage is therefore a property of the estimator, not of the particular value the range probability happens to take in this geometry.

At the largest Grover powers the likelihood in~\eqref{eq:mlqae} develops closely spaced local maxima, because $\sin^{2}((2k+1)\theta)$ oscillates rapidly in $\theta$, and with 100 shots per point the estimate could in principle lock onto a neighbouring branch. Two diagnostics at the largest schedule ($k_{\max}=32$, 40 further trials) test whether that, and not shot noise, limits the error there. A tenfold finer likelihood grid leaves the median error unchanged ($4.5\times10^{-4}$ in both cases), and four times as many shots per point reduce it by a factor of 2.4, close to the factor of 2 expected if shot noise dominates. The high-$k$ error is therefore statistical, and the difference between the two independent sets of 40 trials ($5.5$ and $4.5\times10^{-4}$) lies within the bootstrap interval of figure~\ref{fig:convergence}.

\subsection{Which gates are passed}
\label{sec:gates-passed}

\paragraph{Gate 1 (output structure).} A full dose map on a $128^{3}$ grid requires an independent amplitude estimation per voxel. At a 1\% target, this is $7.0\times10^{9}$ oracle calls against the $10^{7}$ classical histories that a measured GPU engine requires for the same uncertainty~\cite{jia2012}, and which fill all voxels at once. That is a factor of 700 against the quantum approach, before any oracle or hardware cost is counted. Restricting the estimation to the 2\% of voxels carrying high dose leaves a factor of 14. Adopting instead the ten-fold more demanding $10^{8}$-history anchor gives 70 for the full map and 1.4 for the high-dose restriction, still no advantage. Gate 1 depends on no hardware parameter, so this conclusion is not sensitive to future progress: full dose maps are not a target for amplitude estimation.

\paragraph{Gate 2 (amortization).} For $K=10$ scalar summary quantities, classical MC scores all of them from one batch of histories while amplitude estimation repeats $K$ times. At a 1\% target the quantum approach is $13.3\times$ worse; the ratio crosses unity only at $\epsilon=6.9\times10^{-4}$ and reaches $6.7\times$ in favour of the quantum approach at $\epsilon=10^{-4}$. Scalar estimands therefore pass gate 2, but only at precisions well beyond the 1--2\% at which MC dose calculation is normally run.

\begin{figure}[htbp]
\centering
\includegraphics[width=\textwidth]{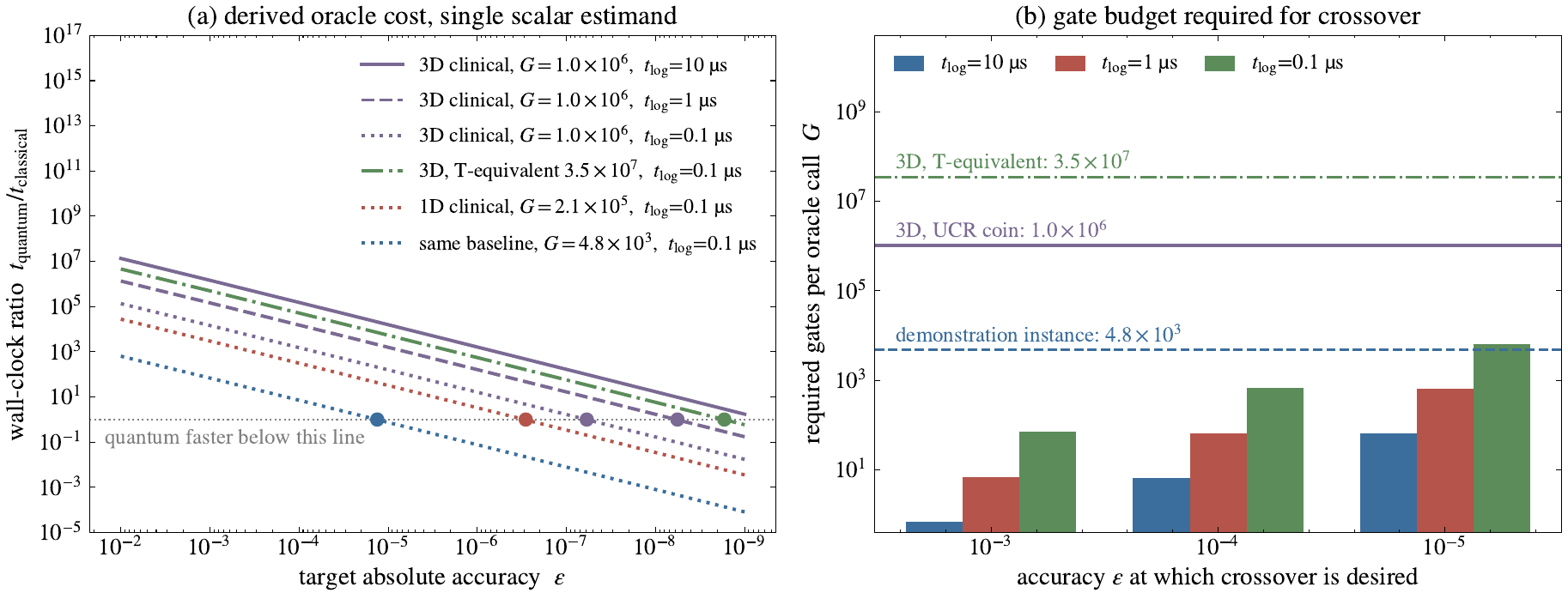}
\caption{Wall-clock analysis with derived oracle gate counts, against a classical baseline of $10^{6}$ histories\,s$^{-1}$. (a) Ratio $t_{\mathrm{quantum}}/t_{\mathrm{classical}}$ for a single scalar estimand, for the clinical-scale oracle with a uniformly controlled coin at three logical-gate times with its $T$-equivalent count at $0.1\,\mu$s, and for the one-dimensional depth-only instance ($G=2.1\times10^{5}$) at $0.1\,\mu$s; markers indicate crossover. All curves, including the one labelled $G=4.8\times10^{3}$, use the same clinical classical baseline, so that curve varies the gate count alone and is not a comparison against the toy problem the demonstration instance solves. (b) Gate budget $G$ required for a crossover at a given target accuracy, with the derived clinical and demonstration gate counts shown for comparison.}
\label{fig:crossover}
\end{figure}

\paragraph{Gate 3 (wall-clock).} The gate count of the oracle, derived in section~\ref{sec:G} rather than assumed, is $4.8\times10^{3}$ logical gates per call for the demonstration circuit and $1.0\times10^{6}$ for a $128^{3}$ clinical geometry with a 10-bit energy register on 415 qubits, or $3.5\times10^{7}$ once the synthesis of its rotations is charged. Figure~\ref{fig:crossover}(a) shows the resulting wall-clock ratio. The crossovers are those of section~\ref{sec:G}: none above $\epsilon=10^{-9}$ at a logical gate time of $10\,\mu$s, $\epsilon\approx6\times10^{-9}$ at $1\,\mu$s, and $\epsilon\approx6\times10^{-8}$ at an aggressive $0.1\,\mu$s, or $1.7\times10^{-9}$ once rotation synthesis is charged. No dosimetric application requires such precisions. The quantum side is charged for a one-dimensional two-outcome model and the classical side for full three-dimensional transport, so these crossovers bound the true distance from below. The remaining curve in figure~\ref{fig:crossover}(a) is the same clinical comparison evaluated at the much smaller gate count of the demonstration instance, $G=4.8\times10^{3}$, where it crosses at $\epsilon\approx1.3\times10^{-5}$; it varies the gate budget alone, as the caption records, and is not a claim about what the 14-qubit demonstration outperforms. What the comparison isolates is that $\epsilon$ moves by more than two orders of magnitude between $G=1.0\times10^{6}$ and $G=4.8\times10^{3}$ at fixed clock and fixed classical baseline, so the gate count, not the estimation method, is what places the crossover out of reach.

Figure~\ref{fig:crossover}(b) inverts the question and asks what gate budget a crossover would require. At $\epsilon=10^{-5}$ and a logical gate time of $0.1\,\mu$s, the most favourable combination considered, the requirement is $G\approx6.4\times10^{3}$. That is approximately the cost of the demonstration instance, a factor of 160 below the clinical-scale gate count and $5.4\times10^{3}$ below its $T$-equivalent count. At $\epsilon=10^{-4}$ with a $1\,\mu$s logical gate the requirement is $G\approx67$, more than four orders of magnitude below. The dominant term in~\eqref{eq:gatecount} is the multiplexed coin, whose cost grows with the number of energy levels; it is also the most compressible, and section~\ref{sec:discussion} examines by how much.

\paragraph{The gates compound.} The three verdicts above are reported at the accuracies where each becomes interesting, but an application must clear all three at the \emph{same} $\epsilon$. Gate 2 turns in favour of amplitude estimation at $\epsilon=6.9\times10^{-4}$; gate 3, at clinical scale, is not cleared anywhere above $\epsilon\approx6\times10^{-8}$. Gate 3 is therefore the binding constraint, and passing gate 2 has no operational consequence while it remains so. Separating them is useful diagnostically, not arithmetically. Gate 1 rules out a problem class outright, gate 2 fixes the precision at which the surviving class could begin to pay, and gate 3 says how far the present circuit is from delivering it.

\section{Discussion}
\label{sec:discussion}

\subsection{What the construction establishes}

Section~1 identified two obstacles. The first was encoding, and the walk addresses it: an energy register and an energy-dependent coin generate a Bragg-shaped depth-deposit profile from transport rules alone, with no precomputed distribution loaded, and the gate count is linear in the number of steps and of energy levels, with no term in $2^{n_{x}}$. The second was the mismatch between amplitude estimation and per-voxel output. We address that by choosing an estimand that is scalar by nature and marked directly in the computational basis, namely the probability that a proton penetrates beyond a distal depth threshold. On this pairing the method works end to end, and the convergence advantage grows with budget as the theory predicts.

What the construction does not establish is a clinical speedup. The wall-clock gate is not passed at clinical scale, and the reason is quantitative, not conceptual: the oracle costs $1.0\times10^{6}$ logical gates and $3.9\times10^{5}$ arbitrary-angle rotations per call, where a crossover at attainable precision would permit of order $10^{3}$ to $10^{4}$ gates. The quadratic scaling of amplitude estimation is a theorem. Whether it can be reached within a clinically meaningful time depends on constants of the kind derived here, and those constants have not previously been computed for a transport oracle in this domain. Nor is sampling the classical method to beat for the model as built (section~\ref{sec:nointerference}): the wall-clock comparison is meaningful for the three-dimensional problem that the clinical register sizes represent, not for the model itself.

\subsection{Where the gate count goes, and how far it can be reduced}

At clinical scale, as a uniformly controlled rotation, the multiplexed coin accounts for 77\% of the per-step gate count, 2048 of 2652 gates, and for every arbitrary-angle rotation in the step, 1024 of them. Implemented literally, as one $n_{E}$-controlled rotation for each of the $L$ energy values, it would cost 17 times as many gates, $4.5\times10^{4}$ per step. Everything else in the step, meaning the two modular subtractors on the energy register and the two controlled adders on the position register, costs 604 gates in total and contains no arbitrary rotation.

The coin is therefore the term worth attacking, and it is compressible because $\theta(e)$ is a smooth function of $e$. Approximating it by $S$ linear segments, each realized as a comparator on the energy register followed by an arithmetic-controlled rotation, gives a cost proportional to $S$ instead of $2^{n_{E}}$. Take comparator and arithmetic-controlled-rotation costs of roughly $8n_{E}$ and $4n_{E}$ two-qubit gates, plus a two-controlled rotation to select the segment. That is 136 gates and 22 rotations per segment at $n_{E}=10$. Against the uniformly controlled coin this changes the gate count little, by factors of 2.3 at $S=4$ and 1.6 at $S=8$, but it cuts the rotations per step from 1024 to 88 and 176, by factors of 12 and 5.8, and the rotations are what error correction charges for. A QROM-style table lookup~\cite{babbush2018} is the obvious alternative but does not help asymptotically, since its cost also scales with the number of table entries.

Coarsening the depth discretization reduces $n_{\mathrm{step}}$ proportionally: 5\,mm bins over the same range give $n_{\mathrm{step}}=154$ instead of 384, and 10\,mm bins give 77. Combining an $S=8$ piecewise coin with 5\,mm bins, for which a 9-bit energy register suffices, yields $G\approx2.4\times10^{5}$ and a $T$-equivalent count of $2.1\times10^{6}$, reductions of 4.2 and 17 from the uniformly controlled construction. That still leaves the gate count $38\times$, and the $T$-equivalent count $320\times$, above the $G\approx6.4\times10^{3}$ a crossover at $\epsilon=10^{-5}$ with a $0.1\,\mu$s logical gate would require. The remaining factor cannot plausibly come from the same source. Once the coin is compressed, the modular arithmetic and the sheer number of transport steps dominate, and neither will give up another order of magnitude without a change to the physics model itself.

Closing the gap therefore requires a different transport formulation, not a better implementation of this one. Multi-step propagators that advance the walker by many depth bins per oracle call, or continuous-variable encodings of the energy loss, would act on $n_{\mathrm{step}}$ instead of on the cost per step, and they seem to us the most promising direction. Whether any such formulation retains the exact unitarity and the direct basis-state marking that make amplitude estimation applicable is an open question.

\subsection{Deposit-weighted metrics and the accumulator}

The estimand used here is a range statistic, an indicator on the final state. Dose-weighted quantities such as the mean dose to a structure, a DVH point or an NTCP input are not indicators. They are expectations of a deposited-energy variable, and a phase oracle on the final state cannot mark them, because the deposit accrues over steps and is not a property of the endpoint.

The standard remedy is an accumulator register: an ancillary register that is incremented by the per-step deposit, conditioned on the coin and energy values, so that the total deposit becomes a computational-basis quantity available at the end of the circuit. Amplitude estimation of $\mathbb{E}[D]$ then follows the value-loading pattern used in quantum finance~\cite{woerner2019} and by Miao \textit{et al}~\cite{miao2026}: a final controlled rotation maps the accumulator value onto the amplitude of a flag qubit, whose measurement probability is the normalized expectation.

We defer it because of the cost. The accumulator adds a controlled modular addition per step, comparable to the existing energy update and therefore modest. But the final value-loading rotation has to be an arithmetic-controlled rotation on the accumulator width, and, more importantly, the accumulator increases the register count and with it the reflection cost inside every Grover iterate. We treat it as the natural next step.

\subsection{Limitations}
\label{sec:limitations}

\paragraph{Physics.} The stopping surrogate $p_{2}(e)$ is not a tabulated stopping power, although replacing it with tabulated values requires no change to the circuit structure. The two-outcome coin gives a Bernoulli energy-loss distribution per step where a Landau-like one is wanted. With losses of one and two units the peak-to-entrance ratio is bounded at 1.690 by~\eqref{eq:sharpbound}, against roughly 3 to 5 in measured curves, and at fixed loss values energy refinement removes straggling as it resolves the energy axis, so the refinement sequence does not converge to a fixed physical model (section~\ref{sec:convergence}). Extending to $K$ loss channels costs $\log_{2}K$ coin qubits per step. The model is one-dimensional and contains neither multiple Coulomb scattering nor nuclear interactions, both of which shape the clinical depth-dose curve.

\paragraph{Simulation scope.} All results are statevector simulations. No hardware execution is reported and no noise model is applied. In the estimation runs the marked-subspace probability is evaluated exactly and the shot counts are then drawn binomially, so the reported errors reflect sampling alone. Gate infidelity, decoherence and state-preparation error would all degrade them, and the fault-tolerant overhead that section~\ref{sec:resource} charges is the price of suppressing those effects. The energy-resolution study at $r\ge2$ uses exact propagation of the chain instead of circuit simulation, which by section~\ref{sec:nointerference} gives the same distribution.

\paragraph{Estimand sensitivity.} Shifting the initial energy by one unit, at fixed $n_{\mathrm{step}}$ and $x_{\mathrm{th}}$, changes the penetration probability from 0.667 to 0.207 at $E_{0}=6$ and to 0.953 at $E_{0}=8$, a central difference of 0.37 per unit. The response is asymmetric because $a$ is bounded by one and reaches it once the whole stopping distribution lies beyond $x_{\mathrm{th}}$, here for $E_{0}\ge9$, so the threshold has to sit within the range distribution for the estimand to be informative. Within that window, the estimand responds strongly to a change in range. That is what makes it a useful range surrogate, and it also means estimator precision is not what limits how finely a range shift could be resolved. We do not convert the achieved precision in $a$ into millimetres: at the present model fidelity that would imply a range resolution the model does not support.

\paragraph{Classical comparator.} The three gates compare amplitude estimation with classical sampling, subject to the precondition G0 (section~\ref{sec:nointerference}). Deterministic transport solvers avoid sampling altogether, and for protons they are already fast: a Boltzmann solver built on Monte Carlo-grade physics computes depth doses and fluence spectra in 5--11\,ms on a single CPU, in agreement with Geant4~\cite{vassiliev2025}. Whether such solvers would also outrun amplitude estimation in three dimensions is a comparison this paper does not make.

\paragraph{Estimator.} The diagnostics of section~\ref{sec:mlqae-results} show the high-$k$ error of the noiseless estimator to be shot-limited. On present hardware a different limit applies first. In a trapped-ion benchmark, maximum-likelihood amplitude estimation saturates at an error of $2$--$5\times10^{-2}$, with its minimum at shallow amplification depth, because depolarization biases the deeper circuits~\cite{lal2026}. Estimators that trade coherent depth for repetitions~\cite{erle2026} are the natural route on early fault-tolerant devices. The gate-3 analysis assumes full error correction and is not affected by either.

\paragraph{Resource model.} The classical anchor and the throughput come from published gPMC timings~\cite{jia2012} and not from a measurement on the reader's own dose engine, and both quantities depend on voxel size, beam energy and hardware generation. Substituting an institution-measured baseline would shift the gate-1 and gate-3 comparisons proportionally, but it would not alter the ordering of the gates: gate 1 fails by between one and three orders of magnitude depending on the anchor, and gate 3 by two to more than three depending on whether rotation synthesis is charged. Gate counts use standard synthesis costs for uniformly controlled rotations and ripple-carry arithmetic. Optimized syntheses would improve the constants, but the coin's rotation count still grows with the number of energy levels, which is the structural point.

\subsection{Outlook}

The three-gate analysis generalizes beyond this construction. Stated as a checklist, a proposal to accelerate a Monte Carlo calculation by amplitude estimation should report:

\begin{enumerate}
\item[(G0)] \textbf{The classical method to beat.} Is sampling the best classical route to the estimand? If the transport model is small enough to be propagated exactly, as the one-dimensional chain of this paper is, neither sampling nor amplitude estimation is needed and G1--G3 are moot. The comparator has to be the best classical method for the problem as posed.
\item[(G1)] \textbf{The estimand and its multiplicity.} What single number is encoded as an amplitude, and how many such numbers $V$ does the application actually require? Amplitude estimation costs $V\,N_{\mathrm{QAE}}(\epsilon)$ where a classical run costs $N_{\mathrm{MC}}(\epsilon)$ once, so a proposal whose output is a field and not a scalar fails here for structural reasons.
\item[(G2)] \textbf{The classical comparator's amortization.} Over how many estimands $K$ does one classical batch amortize, and at what target accuracy $\epsilon$ does $N_{\mathrm{MC}}(\epsilon)$ exceed $K\,N_{\mathrm{QAE}}(\epsilon)$? This locates the precision at which an advantage begins, which is generally far beyond the precision the application needs.
\item[(G3)] \textbf{The oracle's cost per call.} What is $G$, the logical gate count of one $\A$, derived from the circuit and not assumed, and what wall-clock ratio does it give against a measured classical baseline under a stated logical-gate time?
\end{enumerate}

\noindent
The ordering matters. G0, G1 and G2 involve no hardware parameter and can be settled on paper, and a proposal that fails any of them cannot be rescued by hardware progress. Only G3 depends on quantities whose future values are unknown. We would suggest reporting resource statements in this area in these terms alongside any asymptotic speedup factor, since the factor on its own leaves out the constants that decide the matter.

For particle therapy specifically, the near-term value of the construction is likely to lie outside dose calculation proper. A transport oracle that is exactly unitary, and whose coin parameters enter as rotation angles, is an object of independent interest. One example is sensitivity analysis with respect to stopping-power uncertainty, a live problem in proton range verification that involves a few scalar outputs and not a dose map. Whether amplitude estimation is the right tool there is a question we have not addressed, but it is the kind of question the present analysis is meant to make answerable.

\section*{Acknowledgements}
We thank the developers of PennyLane~\cite{pennylane} and the wider open-source scientific Python community.

\section*{Funding}
This research received no specific grant from any funding agency in the public, commercial or not-for-profit sectors.

\section*{Conflicts of interest}
The author declares no conflicts of interest.

\section*{Data availability}
The code implementing the transport oracle, the amplitude-estimation runs, the resource analysis and the figures, together with a driver script and an automated check reproducing every quantity quoted in this paper, is available from the author on reasonable request. No experimental or patient data were used in this study.

\clearpage
\section*{Figures}

\end{document}